\documentclass[usenatbib]{mn2e}

\usepackage{amsmath}
\usepackage{natbib}
\usepackage{epsfig}
\usepackage{txfonts}

\newcommand{\GHz}{{\rm GHz}}

\newcommand{\expf}[1]{{{\rm e}^{#1}}}

\newcommand{\TCMB}{T_{\rm CMB}}

\newcommand{\zs}{{z_{\rm s}}}

\newcommand{\id}{{\,\rm d}}

\newcommand{\beq}{\begin{equation}}   %

\newcommand{\eeq}{\end{equation}}   %

\newcommand{\beqa}{\begin{eqnarray}}   %

\newcommand{\eeqa}{\end{eqnarray}}   %

\newcommand{\beal}{\begin{align}}

\newcommand{\enal}{\end{align}}

\newcommand{\bspl}{\begin{split}}

\newcommand{\espl}{\end{split}}

\newcommand{\bsub}{\begin{subequations}}

\newcommand{\esub}{\end{subequations}}

\newcommand{\bmulti}{\begin{multline}}   %

\newcommand{\beqm}{\begin{mathletters}}   %

\newcommand{\eeqm}{\end{mathletters}}   %

\newcommand{\me}{m_{\rm e}}

\newcommand{\Ne}{N_{\rm e}}

\newcommand{\sigT}{\sigma_{\rm T}}

\newcommand{\vek} [1]{\mbox{\boldmath${#1}$\unboldmath}}

\newcommand{\pd}{\partial}

\newcommand{\pAb}[2]{\frac{\displaystyle\pd #1}{\displaystyle\pd #2}}

\newcommand{\Abl}[2]{\frac{{\rm d} #1}{{\rm d} #2}}

\newcommand{\pot}[2]{#1 \times 10^{#2}}

\usepackage{color}

\usepackage{hyperref}
\usepackage{grffile}
\usepackage{graphics}
\usepackage{booktabs}

\title[Temperature scaling]
{Tests of the CMB temperature-redshift relation, CMB spectral distortions and why adiabatic photon production is hard}

\author[Chluba]{
 J.~Chluba$^{1}$\thanks{E-mail: jchluba@pha.jhu.edu}
 \\
$^{1}$ Department of Physics and Astronomy, Johns Hopkins University, Bloomberg Center 435, 
3400 N. Charles St, Baltimore, MD 21218, USA
}

\date{{\vspace{-3mm}Accepted 2014 June 20. Received 2014 April 28.}}

\begin{document}

\maketitle

\begin{abstract}
In the expanding Universe, the average temperature of the cosmic microwave background (CMB) is expected to depend like $\TCMB\propto (1+z)$ on redshift $z$. Adiabatic photon production (or destruction) or deviations from isotropy and homogeneity could modify this scaling and several observational tests have been carried out in response. Here, we explain why `adiabatic' conditions are extremely difficult to establish in the redshift range targeted by these tests. Thus, instead of leading to a simple rescaling of the CMB temperature, a spectral distortion should be produced, which can be constrained using COBE/FIRAS. For scenarios with late photon production, tests of the temperature-redshift relation (TRR) should therefore be reinterpreted as weak spectral distortion limits, directly probing the energy dependence of the photon production process. 
For inhomogeneous cosmologies, an average $y$-type distortion is produced, but this type of distortion can be created in several other ways. Here, we briefly discuss possible effects that may help disentangling different contributions to the distortion signal, finding this to be very challenging.
We furthermore argue that tests of the TRR, using the SZ effect have limited applicability and that for non-gravitational changes to the TRR the CMB anisotropy spectrum should exhibit an additional $y$-type dependence. 
\end{abstract}



\section{Introduction}
\label{sec:intro}
In the standard Friedmann-Lema\^itre-Robertson-Walker (FLRW) cosmology, the cosmic microwave background (CMB) temperature depends like $\TCMB(z)=T_0(1+z)$ on redshift, $T_0=(2.726\pm0.001)\,{\rm K}$ being the average CMB temperature today \citep{Mather1994, Fixsen1996, Fixsen2009}. Deviations from this scaling would hint towards non-standard cosmological models. Examples include cosmologies with decaying vacuum energy density \citep{Freese1987, Overduin1993, Lima1996, Lima2000} or inhomogeneous cosmologies \citep[e.g.,][]{Losecco2001, Caldwell2008, Zibin2011, Zibin2011b, Marra2011, Bull2013, Caldwell2013}. 

In the first scenario, decay of vacuum energy  leads to `adiabatic' photon production (or destruction), such that the CMB temperature scales like $\TCMB(z)\propto (1+z)^{1-\beta}$ with $\beta\neq 0$. For positive $\beta$, the CMB temperature was slightly lower at higher redshifts than expected from $T_z=T_0(1+z)$, consistent with net photon production until today, while $\beta<0$ implies photon destruction. Generally, $\beta$ can also depend on redshift.
Here, `adiabatic' refers to the fact that when transferring $\Delta \rho_\gamma/\rho_\gamma \ll 1$ of energy to the CMB ($\rho_\gamma$ denoting the CMB blackbody energy density), also a number $\Delta N_\gamma/N_\gamma \simeq (3/4) \Delta \rho_\gamma/\rho_\gamma$ of photons is added \citep{Lima1996, Lima2000}.  As we explain here, this condition alone does not guarantee conservation of the blackbody shape.

The second scenario envisions that we reside in a special location (e.g., a large void) which around us appears isotropic on large scales \citep[see,][for overview]{Goodman1995, Ellis2011}. In other parts of the Universe a deviant average CMB temperature and CMB anisotropies could be seen, e.g., due to anisotropic redshifting or non-standard space-times, violating the {\it Copernican principle} at large scales (i.e., beyond the expected small-scale inhomogeneity even in the FLRW model). This can be used to explain the accelerated expansion of our Universe \citep[e.g.,][]{Alnes2006, Caldwell2008, Ellis2011}, although by requiring very large effective peculiar velocities (${\varv}\simeq 10^4 {\rm km/s}$) radially away from us, ruling out a void model as viable explanation for dark energy \citep[e.g.,][]{GarciaBellido2008, Zhang2011, Zibin2011b, Bull2012} through the kinetic Sunyaev-Zeldovich (SZ) effect \citep{Sunyaev1980}. Still, inhomogeneous models have not been disproven down to the FLRW level, and it is important to consider the limitations of different tests.

One simple way of testing the CMB temperature-redshift relation (TRR) is by looking at the excitation states of interstellar molecules (e.g., C, CO, CN) at various redshifts $z$ \citep{Roth1993, Songaila1994, Noterdaeme2011, Muller2013}. These measurements come with several caveats and should be taken as upper limits on the local CMB temperature, since the pumping of transitions is influenced by many processes \citep[e.g., collisions,][]{Ge1997, Srianand2000}. These tests are mainly sensitive to the average CMB intensity $I_\nu$ in distinct frequency bands, corresponding to the transition energies of low-energy interstellar lines. The current best limits are derived from a combination of measurement of various molecular species $\beta=0.009\pm 0.019$ \citep{Muller2013}, which is consistent with the standard FLRW cosmology at the level $\Delta T/T \lesssim \pot{\rm few}{-2}$. This is still quite a bit above the $\Delta T/T \simeq 10^{-5}-10^{-4}$ level expected in an FLRW Universe, but future measurements with E-ELT\footnote{\url{http://www.eso.org/sci/facilities/eelt/}} or ALMA\footnote{\url{http://www.almaobservatory.org/}} could dramatically improve this (Martins, private communication). 

In addition, modifications of the SZ effect \citep{Zeldovich1969} have been utilized to constrain the TRR \citep{Fabbri1978, Rephaeli1980, Battistelli2002, Horellou2005, Luzzi2009, deMartino2012, Saro2013, Hurier2014}, yielding $\beta=0.009\pm 0.017$ \citep{Hurier2014} from the presently largest cluster sample \citep{Planck2013SZ}. 
The idea of this probe is pretty simple \citep[e.g., see][]{Battistelli2002}. If the effective CMB temperature at the cluster location deviates from $T_z=T_0(1+z)$, the frequency variable $x=h\nu/k \TCMB=h\nu (1+z)^\beta/ k T_z $ is no longer independent of redshift. Therefore, the SZ signal, $S(\nu, p)$, computed with $\beta=0$ and given cluster parameters $p$ (e.g., cluster temperature, bulk velocity and Thomson optical depth), has to be evaluated at frequency $\nu'=\nu(1+z)^{\beta}$ for $\beta\neq 0$. To lowest order in $\beta$, the correction thus depends on the frequency derivative of the cluster signal, $\Delta I_\nu \simeq \nu \,\partial_\nu S(\nu, p) \, \beta \ln (1+z)$ for $|\beta|\ll 1$, where $S(\nu, p)$ can be computed precisely, for example by using {\sc SZpack} \citep{ChlubaSZpack}, also including modifications caused by the line of sight variance of the electron temperature and bulk velocity \citep{Chluba2012moments}. This simple frequency rescaling is degenerate with the modification caused by our own motion with respect to the cluster \citep{Chluba2005b}, an effect that can be easily taken into account due to its known dipolar dependence on the cluster position.

In contrast to these rather local ($z\lesssim 3$) measurements, the CMB temperature anisotropies themselves constrain the possible amount of entropy production which could have happened between the recombination era ($z\simeq 10^3$) and today \citep[e.g.,][]{Opher2004, Opher2005, Borges2008}. Too much photon production since the big bang nucleosynthesis epoch is also unacceptable from measurements of the light element abundances \citep[e.g.,][]{Simha2008, Nollett2011, Jeong2014}.

While it is certainly important to confront any aspect of our cosmology with real observational evidence, here, we wish to highlight that at low redshifts, the aforementioned scenarios lead to a CMB {\it spectral distortion} instead of just a simple change of the CMB temperature. For inhomogeneous cosmologies, the local CMB photon field is given by the superposition of blackbodies with different temperatures \citep[e.g.,][]{Zeldovich1972, Chluba2004, Stebbins2007}. 
This means that the average CMB photon field seen by interstellar molecules or free electrons residing inside clusters generally exhibits a $y$-distortion \citep{Zeldovich1969} with effective $y$-parameter $y\simeq \frac{1}{2}(\Delta T/T)^2$ in addition to a temperature shift. The spectral distortion measurements $\Delta I_\nu/I_\nu \simeq 10^{-4}$ from COBE/FIRAS \citep{Mather1994, Fixsen1996} can therefore be used to place bounds on this scenario \citep[e.g.,][]{Zibin2011, Zibin2011b, Bull2013, Caldwell2013}.

In the future, distortion constraints could further improve to a level $\Delta I_\nu/I_\nu \simeq 10^{-9}-10^{-8}$ with CMB spectrometer concepts similar to PIXIE \citep{Kogut2011PIXIE} or PRISM \citep{PRISM2013WPII}. This suggests that tests of inhomogeneous cosmologies down to the $\Delta T/T \simeq 10^{-4}$ level are within reach; however, early energy release and the reionization and structure formation process could give rise to much larger $y$-distortions \citep[][for recent overview]{Chluba2011therm, Sunyaev2013, Chluba2013fore}. This will compromise our ability to draw precise conclusions from measurements of the average $y$-distortion. Here, we briefly discuss potential ways to disentangle the contributions of different effects, finding this to be very challenging (see Sect.~\ref{sec:inhom_cos_sec}).

For the decaying vacuum scenario (or general models with late photon production/destruction), the argument simply is that the thermalization process \citep{Sunyaev1970mu, Danese1982, Burigana1991, Hu1993} is too inefficient to avoid creating a distortion unless a very finely tuned conversion of the vacuum energy to photons is invoked. Explicitly this means that `adiabatic' conditions are extremely hard to establish and a spectral distortion is practically inevitable (see Sect.~\ref{sec:dec_models}). Indeed, some of these aspects have been alluded to in the original works by \citet{Freese1987}, but here we provide additional clarification. In particular, we explain that with CMB spectral distortion measurements, one could directly probe the energy dependence of the photon production process. We also explicitly show which conditions are required to conserve the blackbody shape at any stage, emphasizing how difficult it is to construct viable models. Finally, we argue that tests of the TRR using the SZ effect have limited applicability (Sect.~\ref{sec:TRR_SZ}) and that for non-gravitational changes to the TRR the CMB anisotropy spectrum should exhibit an additional $y$-type dependence (Sect.~\ref{sec:y_aniso}).

\section{Late photon production or destruction}
\label{sec:dec_models}
One motivation for looking at deviations for the standard redshift dependence of the CMB is driven by the works of \citet{Lima1996} and \citet{Lima2000}, which consider decaying vacuum energy directly coupling to the photon field, leading to `adiabatic' photon production. While it is correct that adding a small number of photons with $\Delta N_\gamma/N_\gamma \simeq (3/4) \Delta \rho_\gamma/\rho_\gamma$ (i.e., constant entropy) does not change the blackbody relations $\rho_\gamma(T)\propto T^4$ and $N_\gamma(T)\propto T^3$, this condition is {\it not} sufficient to avoid creation of a spectral distortion. Just envision that all photons are added in a narrow line at $h\nu/k T \simeq 3.6$. This will do the job \citep[e.g.,][]{Hu1995PhD} but certainly not give a blackbody spectrum unless the thermalization process \citep{Sunyaev1970mu, Danese1982, Burigana1991, Hu1993, Chluba2011therm, Khatri2012b, Chluba2014} is still extremely efficient. 

Thermalization requires both efficient redistribution of photons over frequency and readjustment of the photon number. This is achieved by the combined action of Compton scattering, double Compton and bremsstrahlung emission; however, at $z\lesssim \pot{2}{6}$ (or a few months after the big bang), these processes become inefficient and a distortion is inevitable. Furthermore, at $z\lesssim 10^4$, even Comptonization is already far too slow to transport photons significantly in energy. In this case, photon production in the CMB frequency domain implies that the energy dependence of the coupling cross section will be directly reflected by the CMB distortion. 

To understand this a bit better, let us consider the evolution of the average photon occupation number, $n_\nu$, which is determined by
\beq\label{eq:BoltzEq_Photons}
\pAb{n_{\nu}}{t}-H\,\nu\,\pAb{n_{\nu}}{\nu}
=\left.\Abl{n_{\nu}}{t}\right|_{\rm C}
+\left.\Abl{n_{\nu}}{t}\right|_{\rm DC}
+\left.\Abl{n_{\nu}}{t}\right|_{\rm BR}
+\left.\Abl{n_{\nu}}{t}\right|_{\rm S}.
\eeq
The second term on the left hand side describes the redshifting of photons due to the adiabatic expansion of the Universe ($H$ denoting the Hubble factor) and the right hand side terms correspond to the physical processes mentioned above. 
We also explicitly added a photon source term, $\left.\id n_{\nu}/\id t\right|_{\rm S}$, envisioning the effect of vacuum energy decay or photon destruction.

For our purposes ($z\lesssim 10^4$), we can neglect the double Compton and bremsstrahlung processes, $\left.\id n_{\nu}/\id t\right|_{\rm DC}\approx \left.\id n_{\nu}/\id t \right|_{\rm BR}\approx 0$. Unless, we are interested in effects in the  far Rayleigh-Jeans part of the spectrum ($h\nu \lesssim 10^{-2}\,k\TCMB$), where late free-free emission and absorption are still efficient \citep{Chluba2011therm}, this should be a very good approximation.
Then, two main regimes are important: if the characteristic energy of photons created by the decay of vacuum energy density (or another processes) is $h\nu \gg 100 \,k \TCMB$ (i.e., these are not really CMB photons), we have $\Delta \rho_\gamma/\rho_\gamma\gg \Delta N_\gamma/N_\gamma$ and the injected photons will quickly heat electrons and baryons \citep{Shull1985, Slatyer2009, Huetsi2009} leading to subsequent up-scattering of CMB photons and creating an average $y$-distortion visible in the CMB \citep[e.g.,][]{Chluba2013Green, Chluba2013fore}. In this case, the Compton scattering term, $\left.\id n_{\nu}/\id t\right|_{\rm C}$, through the interaction with the slightly heated thermal electrons, sources the distortion with all CMB photons simultaneously up-scattering only slightly\footnote{We neglect the possible extra emission of secondary soft photons from very high energy events ($h\nu > 511\,{\rm keV}$), which would give rise to a particle cascade and non-thermal pairs.}. 
Here, the limit $|y|\lesssim \pot{1.5}{-5}$ from COBE/FIRAS \citep{Fixsen1996} provides the tightest constraint, implying that at most $\Delta \rho_\gamma/\rho_\gamma \simeq \pot{6}{-5}$ of energy was transferred to the CMB in the $y$-distortion era at $z\lesssim 10^4$.
For $\simeq {\rm keV}-{\rm TeV}$ energies, the low-redshift Universe furthermore becomes practically transparent to photons \citep{Chen2004}, so that these scenarios can be directly constrained by measurements of the diffuse X-ray and $\gamma$-ray background \citep{Atwood2009, Abdo2010, Zavala2011}.

If on the other hand, the characteristic energy of the produced photons is $h\nu \lesssim 100 \,k \TCMB$, then the Compton scattering process only leads to a small change of the emitted photon distribution but hardly any extra energy exchange with the electrons. At low frequencies ($h\nu\lesssim 0.5 k \TCMB$), the associated scattering-induced correction can be captured using analytic solutions that include the effect of stimulated scattering in the CMB ambient radiation field \citep{Chluba2008d}; at $0.5 k \TCMB\lesssim h\nu\lesssim 10 k \TCMB$, stimulated effects can be neglected and the classical solution of \citet{Zeldovich1969} is applicable; however, at late times ($z\lesssim 10^3$), these modifications are small. At higher frequencies, the recoil effect also becomes noticeable, leading to a drift of photons towards lower frequencies with characteristic shift $\Delta \nu / \nu \simeq h\nu/\me c^2$ per scattering. Still, at $z\lesssim10^4$, even recoil is rather inefficient \citep[at $z\lesssim 7000$ and $h\nu \lesssim 40 \,k \TCMB$, the total recoil until today is $\Delta \nu/\nu\lesssim 7\%$;][]{Jose2008} so that photons are mainly stuck where they are and just redshift until they, for example, can re-excite and ionize hydrogen and helium atoms in the recombination era \citep{Chluba2007b, Chluba2009c, Chluba2010a} or destroy H$_2$ molecules at much later time \citep{Carla2013}. Thus, the solution for the spectral distortion in this regime is roughly
\beq\label{eq:BoltzEq_Photons_sol}
\Delta n(x, z)\approx  \int_z^{\zs} \left.\Abl{n(x, z')}{z'}\right|_{\rm S} \id z',
\eeq
where $\zs\lesssim 10^4$ (following our approximations) is the redshift at which the photon production or destruction began. We also transformed the photon source term to redshift and dimensionless frequency $x=h\nu / k T_z \equiv {\rm const}$ with $T_z=T_0(1+z)$. 

To ensure that the blackbody shape is not affected by the photon production process, we need
\beq\label{eq:no_dist}
\left.\Abl{n(x, z)}{z}\right|_{\rm S}\approx \frac{x\expf{x}}{(\expf{x}-1)^2} \, \phi(z),
\eeq
which implies a very special energy dependence for the coupling between vacuum energy and photons. In particular, it has to be directly related to the temperature derivative of the CMB blackbody at a given moment. Here, we assumed $|\int \phi(z')\id z'| \ll 1$, requiring that the total change of the CMB temperature, $\Delta T/T \approx \int_z^{\zs} \phi(z')\id z'$, remains small (such that second-order terms can be neglected). 
Assuming that $\TCMB(z)=T_0(1+z)^{1-\beta}\equiv T_z (1+z)^{-\beta}$ with constant $\beta$ is created, we find $\phi(z)\approx \beta/(1+z)$ and thus $\Delta T/T \approx  \beta \,[\ln(1+\zs)- \ln (1+z)]$ with respect to the initial temperature at $\zs$. Normalizing instead to the CMB temperature today, we similarly have $\Delta T/T \approx  -\beta \ln (1+z)$ and $\phi(z)\approx -\beta/(1+z)$.
If the energy dependence of the photon source term deviates from Eq.~\eqref{eq:no_dist}, a spectral distortion is inevitable and these relations are not applicable. This condition is much more stringent than what is usually referred to as `adiabatic' photon production.

The above analysis demonstrates that measurements of the CMB temperature at higher redshifts generally do not provide any improved limit on photon production scenarios. Generically, these scenarios should lead to a spectral distortion and `adiabatic' photon production is a very finely tuned case. According to Eq.~\eqref{eq:BoltzEq_Photons_sol}, the limits on deviations from the blackbody shape obtained with COBE/FIRAS at frequencies $60\,\GHz\lesssim \nu \lesssim 630\,\GHz $ \citep{Fixsen1996} can thus be used to directly constrain the energy dependence of the coupling to photons on a case-by-case basis. Measurements of different excitations states at higher redshift should thus be reinterpreted as additional limits on CMB spectral distortions. However, given observational uncertainties, these bounds should presently be some two orders of magnitudes weaker than those from COBE/FIRAS.
Scenarios with late photon production due to photon-axion \citep{Tashiro2013, Ejlli2013, Conlon2014}, photon-graviton \citep{Dolgov2013, Ejlli2013b} and disformal coupling \citep{Brax2013, Bruck2013} were also discussed in the literature. Generally, these scenarios all produce CMB spectral distortions. 
%

\vspace{-4mm}
\section{Inhomogeneous cosmologies}
\label{sec:inhom_cos_sec}
In the previous section, we assumed that the photon production is uniform and that the background cosmology is homogenous and isotropic at the largest scales. Non-uniform photon production would lead to spatial variations of the CMB spectrum, but again `adiabatic' conditions are hard to achieve unless local physics creates photons with an energy distribution that is extremely close to the one of Eq.~\eqref{eq:no_dist}. Still, measurements of the excitation states of different interstellar species and searches for spectral features in the SZ signal could in principle be used to derive limits on spatial variations of the CMB spectrum. Also, COBE/FIRAS and in the future measurements with a PIXIE-type spectrometer can provide stringent constraints at least on the largest scales.

Next, let us consider inhomogeneous cosmologies without any extra photon production. Observations of the CMB anisotropies \citep{WMAP_params, Planck2013params} and large-scale structure \citep[e.g.,][]{Tegmark2004,Heymans2012} imply a very high degree of isotropy at the largest scales, at least with respect to our own location. One way to change the TRR is by extra gravitational red- and blueshifting around large mass concentrations. This possibility is strongly constrained by measurements of the galaxy distribution and gravitational lensing; so that even in extreme cases, local temperature corrections larger than $\Delta T/T\simeq 10^{-4}$ seem unlikely and well below present observational capabilities \citep[see also discussion in][]{Losecco2001}. While measurements of the excitation state of interstellar species in principle are sensitive to these effects, the SZ effect will {\it not} be affected, because to leading order the net gravitational effect cancels once the scattered photons climbed back out of the potential well to reach us. 

Another way to create a departure from the standard TRR is by {\it anisotropic redshifting} (similar to above but on larger scales). Since from our location, the Universe appears largely isotropic, this scenario is limited to radially inhomogeneous cosmologies (i.e., close to spherically symmetric around us). These models are tightly constrained by measurements of the $y$-parameter from COBE/FIRAS \citep{Zibin2011, Zibin2011b, Caldwell2013} and the kinetic SZ (kSZ) effect \citep{Moss2011, Zhang2011, Bull2012, Planck2014kSZ}. Both tests are mainly sensitive to the dipole anisotropy induced for inhomogeneous cosmology models. 
The physics again is very simple: the local CMB field is given by a superposition of blackbodies of different temperatures. The temperature differences are created due to anisotropies in the effective redshifting rate and the average local spectrum is thus no longer a simple blackbody. Free electrons can then deflect photons in and out of our direction, giving us notice of the large anisotropies that are otherwise hidden from us \citep{Goodman1995}. Different models predict different amplitudes for the CMB anisotropies and changes to the local monopole which then allows constraining them.

Denoting the temperature field in different directions $\hat \gamma$ around some location $\vek{x}$ as $T(\vek{x}, \hat\gamma)=T_z+\Delta T(\vek{x}, \hat\gamma)$ with $T_z=T_0(1+z)$ and $\Delta T/T_z\ll 1$, the local CMB photon occupation number is given by \citep[e.g.,][]{Chluba2004, Stebbins2007}
\beal\label{eq:sup_BB}
n_\nu(\vek{x}, \hat\gamma)
&\approx \frac{1}{\expf{x}-1}+G(x)\,\left[\Theta(\vek{x}, \hat\gamma)+\Theta^2(\vek{x}, \hat\gamma)\right]+\frac{1}{2} Y_{\rm SZ}(x) \, \Theta^2(\vek{x}, \hat\gamma)
\nonumber
\\[0mm]
&= \frac{1}{\expf{x}-1}+G(x)\,\Theta(\vek{x}, \hat\gamma)+\frac{1}{2} Y^\ast_{\rm SZ}(x)\, \Theta^2(\vek{x}, \hat\gamma)
\end{align}
with $\Theta(\vek{x}, \hat\gamma)=\Delta T(\vek{x}, \hat\gamma)/T_z$. Here, we defined the spectral functions $G(x)=-x\partial_x (\expf{x}-1)^{-1}= x\expf{x}/(\expf{x}-1)^2$ for a simple temperature shift, $Y_{\rm SZ}(x)=G(x)\left[x\coth(x/2)-4\right]$ for a $y$-distortion and $Y^\ast_{\rm SZ}(x)=G(x)\left[x\coth(x/2)-2\right]$ for a $y$-type distortion that vanishes in the limit $x\ll 1$ but scales as $Y^\ast_{\rm SZ}(x)\approx Y_{\rm SZ}(x)$ for $x\gg 1$.

We shall assume spherical symmetry around us, so that we can write $\Theta(\vek{x}, \hat\gamma)=\sum_{\ell=0} (2\ell +1) \Theta_\ell(z) P_\ell(\hat\gamma'\cdot \hat \gamma)$, where $P_\ell(x)$ denote Legendre functions, $\hat \gamma'$ defines our line of sight and $z$ the redshift of the location $\vek{x}$. Integrating over all directions, we thus have $\left<n_\nu\right>\approx 1/(\expf{x}-1)+G(x) \left<\Theta\right>+\frac{1}{2} Y^\ast_{\rm SZ}(x)\, \left<\Theta^2\right>$ for the average CMB photon occupation number. Here, $\left<\Theta\right>=\Theta_0(z)$ and $\left<\Theta^2\right>=\sum_{\ell=0} (2\ell+1)\Theta_\ell^2(z)$, such that one can also write
\beal\label{eq:sup_BB_alt}
\left<n_\nu\right>\approx \frac{1}{\expf{x'}-1}+\frac{1}{2} Y^\ast_{\rm SZ}(x)\, \sum_{\ell=1} (2\ell+1)\Theta_\ell^2(z)
\end{align}
with $x'=x/[1+\Theta_0(z)]$. The first term describes a blackbody at temperature $T'=T_z[1+\Theta_0(z)]$, which for small $\Theta_0(z)$ translates into $\beta(z) \simeq -\Theta_0(z)/\ln(1+z)$. The second term describes the distortion part. Usually this is a negligible correction; however, in the Wien tail of the blackbody spectrum, it could become significant, since $Y^\ast_{\rm SZ}(x)/G(x)\simeq x$. 
Assuming that $\Theta^2_0(z)\simeq \sum_{\ell=1} (2\ell+1)\Theta^2_\ell(z)$, this suggests that only at high frequencies, $x\gtrsim 2/\Theta_0(z)\gtrsim 100$, the distortion becomes large, and otherwise the distortion part only causes a correction, confirming our statement. Thus, measurements of the excitation states of different interstellar species, which are mainly sensitive at lower frequencies, do not allow constraining the distortion part. In addition, the cosmic infrared background strongly contaminates the Wien tail of the CMB.

\vspace{-3mm}
\subsection{Scattering signals in the Thomson limit}
\label{sec:kSZ_tau}
To check the consistency of the model, one would in principle like to measure $\Theta_\ell(z)$ for $\ell >0$. Since it is hard to achieve the precision required to constrain the distortion term in Eq.~\eqref{eq:sup_BB_alt} using interstellar medium measurements, scattering signals could provide an alternative \citep{Goodman1995}. In the Thomson limit (no energy exchange between photons and scattering electron at rest), only the monopole and quadrupole anisotropy of the local radiation field scatters back into the line of sight. For the monopole, the net scattering effect cancels, while for the quadrupole, a non-zero net effect remains. For all other multipoles of the radiation field, scatters act as an {\it optical depth screen}, just removing photons from the line of sight. 

We assume that the CMB anisotropies $\Theta_\ell(z)$ are those `seen' by a scatterer \citep[e.g., see][for more explanations]{Zibin2011b}.
With Eq.~\eqref{eq:sup_BB_alt}, we can rewrite Eq.~\eqref{eq:sup_BB} as
\beal\label{eq:sup_BB_anis}
n_\nu(z, \hat\gamma)
&\approx \left<n_\nu\right>+
G(x)\,\sum_{\ell=1} (2\ell+1)\Theta_\ell(z) P_\ell(\hat\gamma'\cdot \hat \gamma)
\nonumber\\[-1mm]
&\qquad\quad +\frac{1}{2} Y^\ast_{\rm SZ}(x)\, 
\sum_{\ell=1} (2\ell+1)\Omega_\ell(z) P_\ell(\hat\gamma'\cdot \hat \gamma),
\end{align}
where $\Omega_\ell(z)=\frac{1}{2}\int [\sum_{\ell'=0} (2\ell'+1)\Theta_{\ell'}(z) P_{\ell'}(x)]^2 P_\ell(x)\id x$ for $\ell>0$ gives the projections of $\Theta^2(\vek{x}, \hat\gamma)$ on to different Legendre functions. We find $\Omega_\ell(z)=\sum_{\ell',\ell''=0}(2\ell'+1)(2\ell''+1)\Theta_{\ell'}(z)\Theta_{\ell''}(z)\,G_{\ell \ell' \ell''}$, where $G_{\ell \ell' \ell''}=\frac{1}{2}\int P_\ell(x) P_{\ell'}(x) P_{\ell''}(x) \id x$ are the Gaunt integrals. In the single scattering limit, the correction to the radiation field thus is
\beal
\Delta n_\nu(\hat\gamma')
&\approx \Delta \tau(\hat\gamma') \left\{\frac{3}{16\pi} \!\!\int \!\!\id^2 \hat\gamma \left[1+(\hat\gamma'\cdot \hat\gamma)^2\right]n_\nu(z, \hat\gamma)-n_\nu(z, \hat\gamma')\right\}
\nonumber\\
&=-G(x)\sum_{\ell=1} (2\ell+1) \Theta_\ell(z) \Delta  \tau(\hat\gamma') 
\nonumber\\[-1mm]
&\qquad - \frac{1}{2} Y^\ast_{\rm SZ}(x) \sum_{\ell=1} (2\ell+1) \Omega_\ell(z)\Delta   \tau(\hat\gamma')
\nonumber\\[-1mm]
&\qquad\qquad + \frac{1}{4} Y^\ast_{\rm SZ}(x) \Omega_2(z)\Delta   \tau(\hat\gamma'),
\label{eq:sup_BB_anis_scat_gen}
\end{align}
where $\Delta \tau = \sigT \Ne \Delta l$ is the differential Thomson optical depth through a small fluid element around redshift $z$. The first two terms in Eq.~\eqref{eq:sup_BB_anis_scat_gen} account for photons scattering out of the line of sight, while the last term gives the quadrupole contribution scattering back into the line of sight.

Simulations of the void scenario typically predict very small amplitudes for multipoles beyond the dipole $\ell >1$ \citep{Alnes2006, GarciaBellido2008}. Assuming that the local radiation field only has a temperature dipole, we have $\Theta_1(z)\neq 0$ and $\Omega_2(z)=(6/5)\Theta^2_1(z)$, yielding
\beal
\label{eq:sup_BB_anis_scat}
\Delta n_\nu(\hat\gamma')
&=-3 G(x) \Theta_1(z) \Delta  \tau(\hat\gamma') -Y^\ast_{\rm SZ}(x) \frac{27}{10} \Theta^2_1(z)\Delta \tau(\hat\gamma').
\end{align}
The first term is just due to the dipolar part of the local radiation field, which only scatters out of the line of sight, while the last term is the net effect from the quadrupolar part. Assuming that the dipole is caused by the motion of the scatters, i.e. ${\varv}/c=-3\Theta_1$, with $G(x)\rightarrow G(x)[1-{\varv}/c]-Y^\ast_{\rm SZ}(x)({\varv}/c)$ [Lorentz transformation into the CMB rest frame] we find
\beal
\label{eq:sup_BB_anis_scat_II}
\Delta n_\nu(\hat\gamma')
&\approx G(x)\int\! 
\frac{{\varv}}{c} \id \tau(\hat\gamma')
+\frac{7}{10}Y^\ast_{\rm SZ}(x) \int\! \left(\frac{\varv}{c}\right)^2 \!\id \tau(\hat\gamma')
\end{align}
after integrating along the line of sight.
The last term is the expression for the distortion, where $y\simeq \frac{7}{10}\int\left(\varv/c\right)^2\!\!\id \tau(\hat\gamma')$ defines the effective $y$-parameter introduced by inhomogeneities \citep{Moss2011}. It was initially derived using the work of \citet{Stebbins2007}, but can also be obtained from second-order Doppler terms of the photon Boltzmann equation \citep[e.g.,][]{Hu1994pert, Nozawa1998}. The limits from COBE/FIRAS then imply $|\varv|/c \lesssim \pot{1.5}{-2}$  for constant $\varv$ and $\tau \simeq 0.1$ from WMAP \citep{Komatsu2010}.

The first term in Eq.~\eqref{eq:sup_BB_anis_scat_II} is related to the kSZ effect caused by the large-scale motion of the scatterers. We neglected perturbations in the electron density and only consider at the average density at different distances from the center. In the void model, all matter moves away from the center, so that this term is negative overall. Interestingly, this gives rise to a small temperature correction, $\Delta T/T_0\simeq \int \!\frac{{\varv}}{c} \id \tau<0$, to the observed value of our CMB monopole which is introduced by scattering\footnote{The term $\simeq \frac{1}{2}Y_{\rm SZ}^\ast(x) \left[\int \frac{{\varv}}{c}\! \id \tau\right]^2$ missing to conserve the blackbody part should arise when including the effect of second scattering.}  \citep[see also,][]{Zibin2011b}. 
At our present location, this term is expected to be smaller than $\Delta T/T \simeq 10^{-4}-10^{-3}$; however, this depends on the model and at $z>0$, the correction could be larger, in principle providing a way to induce $\Theta_0\neq 0$ and $\beta\neq 0$. This scenario could be probed using the excitation state of interstellar species and the SZ effect (see below).

\subsection{Can we disentangle different distortion contributions?}
The reionization process heats the medium to an average temperature $T\simeq 10^4\,{\rm K}$. This introduces a global $y$-distortion $y\simeq \pot{\rm few}{-7}$ \citep{Sunyaev1972b, Hu1994pert}. Shocks and supernova explosions can further heat the medium to $T\simeq 10^5\,{\rm K}$, pushing the average $y$-parameter to a level $y\simeq 10^{-6}$ \citep{Cen1999, Refregier2000, Oh2003, Zhang2004}. The peculiar motions of the gas also induce a second-order Doppler distortion, with effective $y$-parameter $y\simeq \frac{1}{3} \int ({\varv}_{\rm p}/c)^2 \id \tau(\hat\gamma')\simeq \pot{\rm few}{-8}$ \citep[e.g.,][]{Hu1994pert, Goodman1995, Stebbins2007}.
Thus, constraints on inhomogeneous cosmologies or conversely limits on the reionization and structure formation process using future CMB spectral distortion measurements ultimately are limited by how well one can disentangle their contributions to the average signal. Similarly, a $y$-distortion produced in the pre-recombination era by energy release (e.g., decaying particles or the dissipation of primordial acoustic modes) could contribute at a significant level.

How do we separate all these cases? It is very hard to give an answer to this question and here we only offer a list of possible directions.
For pre-recombination $y$-distortions, a correction to the cosmological recombination spectrum \citep{Sunyaev2009} is expected, which might help distinguishing it from post-recombination $y$-distortions \citep{Chluba2008c}. This does require very precise CMB spectroscopy, but for a large pre-recombination $y$-distortion ($y\simeq 10^{-5}$ in agreement with COBE/FIRAS), due to an enhancement of the recombination radiation by roughly two orders of magnitude, this effect may already be visible with a PIXIE-type experiment.

Separating the reionization and structure formation distortion from a signal induced by inhomogeneous cosmologies is harder. As pointed out above, one effect is a scattering-induced change to the local CMB monopole, which leads to a small modification of the TRR. This process mimics the effect of adiabatic photon production but is just related to the spatial redistribution of photons by electron scattering in and out of our line of sight. Generally, the modification can be non-monotonic and gives a correction to the TRR in addition to the Sachs-Wolfe effects, imply $\Theta_0(z)\neq 0$. Both effects can in principle be constrained using interstellar species, but the SZ effect only works for the scattering-induced correction (see below). At the level $y\simeq 10^{-6}$, one furthermore does not expect $|\varv|/c$ to be much in excess of $|\varv|/c\simeq 10^{-3}/\sqrt{0.7 \tau}\simeq \pot{4}{-3}$ with $\tau\simeq 0.1$, and thus, $\Theta_0(z)\lesssim \pot{\rm few}{-4}$ on average. This shows that very high precision is required, but this would provide one important consistency check for inhomogeneous cosmology models.

Another potential way to disentangle different effects could be to look at the scattering of CMB photons including the energy exchange with the hot cluster electrons. This will not only affect the local CMB monopole, but at lowest order in the electron temperature also leads to a small scattering correction from the dipole, quadrupole and octupole parts of the radiation field, changing their spectra \citep{Chluba2014mSZI, Chluba2014mSZII}. If the temperature anisotropies are much larger than expected in an FLRW cosmology, this may provide a way to directly constrain $\Theta_\ell(z)$ for $\ell \leq 3$. For a single cluster with Compton $y$-parameter $y\simeq 10^{-4}$, the effect is $\simeq y\, \Theta_\ell$, which only is a small correction to the thermal SZ effect.
One problem is that the second scattering correction to the cluster SZ signal itself can be comparable in amplitude, though with slightly different frequency dependence \citep{Chluba2014mSZI, Chluba2014mSZII}.
Similarly, differences in the cluster atmospheres [such as relativistic temperature corrections \citep{Sazonov1998, Itoh98, Challinor1998} and line of sight electron temperature variations \citep{Chluba2012moments}] have to be carefully understood. Also, the correction caused by our own motion with respect to the cluster \citep{Chluba2005b} and non-thermal SZ contributions \citep{Ensslin2000, Colafrancesco2003} have to be taken into account.
All these effects related to SZ clusters make it very hard to separate different terms.

Finally, we mention that scattering of the quadrupole anisotropy could produce a weak polarization signal $\simeq 0.1 \tau \Theta_{2,\pm 1}$, where $\Theta_{\ell m}$ denotes the spherical harmonic coefficients of the temperature field with respect to our line of sight. By looking at the large-scale CMB E-mode polarization anisotropies or SZ cluster polarization \citep{Kamionkowski1997}, one could therefore also constrain a possible non-standard redshift scaling of the CMB quadrupole. This would provide a useful consistency check, however, for radial symmetry (i.e., $n_\nu$ has azimuthal symmetry around $\hat\gamma'$ and $\Theta_{\ell m }=\delta_{m0} \Theta_{\ell 0}$), no polarization signal is produced in the single scattering limit \citep[see, ][for more discussion of polarization effects]{Sazonov1999}.

\vspace{-4mm}
\section{When do SZ tests of the TRR work?}
\label{sec:TRR_SZ}
We already mentioned that tests of the TRR using clusters have limited applicability. First of all, if the local CMB temperature is changed by gravitational effects, then to lowest order the SZ signal will not be sensitive to this, since both the CMB blackbody part and the SZ signal acquire the same net redshifting effect while traveling through the potential landscape on the way towards us. This conclusion also applies if a the potentials vary in time, causing an integrated Sachs-Wolfe effect, as the relative position and shapes of the two spectral parts remains the same. 

We also argued that uniform photon production (by some exotic source term) generally should cause a spectral distortion and thus is tightly constrained by COBE/FIRAS, making a simple change in the TRR by this process very unlikely. However, if we still assume that photons were indeed added without changing the CMB spectrum, then the SZ cluster signal allows probing the change in the CMB temperature between the cluster redshift and today (see Introduction).
Here, the crucial point is that the CMB blackbody part has to be changed without simultaneously affecting the SZ signal itself, such that the two spectra are shifted with respect to each other.
Thus, net changes in the TRR caused by a global kSZ effect from inhomogeneous cosmologies (Sect.~\ref{sec:kSZ_tau}) could potentially also be constrained using SZ cluster samples.
The additional correction to the SZ signal introduced by scatters between the cluster and us is, $\Delta S(x)\approx -S(x) \int \id \tau$. This means that the total SZ signal is slightly suppressed in amplitude with respect to the case without scattering, $S_{\rm tot}(x)\approx [1-\tau(z, 0)]S(x)$, where $\tau(z,0)=\int^z_0 \id \tau$. This is degenerate with the optical depth of the cluster itself and does not change the shape of the SZ signal. On the other hand, the kSZ effect from the motion of these scatters modifies the TRR, so that the SZ signal is shifted with respect to the CMB blackbody part. This test only works if the net change in the CMB temperature from the cluster redshift to $z=0$ is non-zero with respect to the standard law, $T_z=T_0(1+z)$. It is not necessary that the change in the TRR is monotonic. It is furthermore strongly model dependent by how much the temperature of the CMB between the cluster redshift and today is affected by the kSZ, so that a more detailed case-by-case study is required. 

\vspace{-4mm}
\section{Distortion of the CMB anisotropies}
\label{sec:y_aniso}
Although we already explained that adiabatic photon production is difficult, here we show that this would also cause $y$-type CMB anisotropies. This provides another way to test non-gravitational changes to the TRR, for instance introduced by a global kinetic SZ effect.
Unless, photon production occurs in a way that also conserves the spectrum of the CMB anisotropies, there are three main effects: (i) the contrast of CMB temperature fluctuations decreases (let us assume photon injection), (ii) the redshift of recombination is higher and (iii) the spectra of the anisotropies are `out of tune' with the higher monopole temperature at $z=0$. 
Effects (i) and (ii) were discussed earlier \citep{Opher2004, Opher2005}, while (iii), to our knowledge, was not mentioned before.

All three effects can be understood when assuming that photon production only sets in after recombination. At $z_{\rm s}< z < z_{\rm re} \simeq 10^3$ (i.e., before photon production but after recombination), the CMB photon occupation number in any direction $\hat \gamma$ around us takes the form $n_\nu(z, \hat\gamma)\approx 1/(\expf{x^\ast}-1)+G(x^\ast)\,\Delta T(z, \hat\gamma)/T^\ast$ with $x^\ast=h\nu/kT^\ast$, but where $T^\ast = T_0^\ast (1+z)$ with $T_0^\ast<T_0$. The CMB anisotropies just evolve in the usual way with slightly modified initial perturbations because of the different initial distance to the last scattering surface, leading to $n_\nu(0, \hat\gamma)\approx 1/(\expf{x^\ast_0}-1)+G(x^\ast_0)\,\Delta T(0, \hat\gamma)/T^\ast_0$ at redshift $z=0$. The power spectrum $C^{TT}_\ell$ of $\Theta^\ast(\hat\gamma)=\Delta T(0, \hat\gamma)/T^\ast_0$ can be obtained from CAMB \citep{CAMB} using $T^\ast_0<T_0$ instead of $T_0$.
Assuming that adiabatic conditions for the photon production were achieved, we furthermore have a correction to the CMB monopole such that the total distribution is given by $n'_\nu(0, \hat\gamma)\approx 1/(\expf{x_0}-1)+G(x^\ast_0)\,\Theta^\ast(\hat\gamma)$, with $x_0=h\nu/kT_0$. This shows that the spectrum of the CMB anisotropies would have a slightly different effective temperature, $T^\ast_0<T_0$. We can rewrite this as
\bsub
\label{eq:sup_BB_mod_anis}
\beal
n'_\nu(0, \hat\gamma)
&\approx \frac{1}{\expf{x_0}-1}+G(x_0 T_0/T_0^\ast)\,\Theta^\ast(\hat\gamma)
\nonumber\\
&\approx \frac{1}{\expf{x_0}-1}+G(x_0)\,\Theta^\ast(\hat\gamma)\left[1+\frac{\Delta T^\ast_0}{T_0}\right]
\nonumber\\[-1mm]
&\qquad\qquad  + Y^\ast_{\rm SZ}(x_0) \, \Theta^\ast(\hat\gamma)\frac{\Delta T^\ast_0}{T_0}  + \mathcal{O}(\Delta T/T)^3
\label{eq:dilution_II}
\end{align}
\esub
to isolate the three effects mentioned above. The anisotropy dilution factor $f=(1+ \Delta T^\ast_0/T_0)\equiv T^\ast_0/T_0<1$ (it can be an enhancement factor if photons are removed) arises from the frequency rescaling $G(x_0 T_0/T_0^\ast)$. Changes in the redshift of recombination are directly encoded by $\Theta^\ast(\hat\gamma)$. The $y$-type distortion of the anisotropy spectrum is given by the last term\footnote{We neglected the second-order terms for the initial photon distribution which are much smaller unless $\Delta T^\ast_0/T_0$ is close to the FLRW limit.}.
Depending on the sign of $\Delta T^\ast_0/T_0$, the distortion correlates/anticorrelates with the temperature anisotropies: $C^{y T}_\ell(x) \simeq (\Delta T^\ast_0/T_0) C^{TT}_\ell [Y^\ast_{\rm SZ}(x)/G(x)]$ with $Y^\ast_{\rm SZ}(x)/G(x)\simeq x$ at high frequencies. The effective $y$-parameter, $\Delta y(\hat\gamma)~\simeq~\Theta^\ast(\hat\gamma)(\Delta T^\ast_0/T_0)$, can furthermore be both positive and negative, depending on the sign of $\Theta^\ast(\hat\gamma)$, with auto-power spectrum $C^{y y}_\ell(x) \simeq (\Delta T^\ast_0/T_0)^2 C^{TT}_\ell [Y^\ast_{\rm SZ}(x)/G(x)]^2$. The effect would be mainly visible in the Wien tail of the CMB, as $Y^\ast_{\rm SZ}(x_0)$ vanishes for $x\ll 1$. 
It is purely due to the fact that the phase space distribution of the temperature anisotropies is frozen in shape. With respect to a CMB monopole $T_0\neq T_0^\ast$, the spectrum of the anisotropies is thus distorted or equivalently has a different effective temperature.

Assuming $\Delta T^\ast_0/T_0\simeq 1\%$, spatially varying $y$-type distortions with amplitude $\Delta y\simeq 10^{-7}$ should be observable with a PIXIE-type experiment. Interestingly, this would allow us to check if our CMB dipole, $\Delta T_{\rm d}/T\simeq 10^{-3}$, indeed is motion induced, since otherwise, we should see a large $y$-type dipole $\Delta y_{\rm d}\simeq 10^{-5}$. With measurements of the dipole spectrum \citep{Fixsen1996}, assuming that the CMB dipole is fully primordial, we find $\Delta T^\ast_0/T_0=\pot{(9.2\pm 4.9)}{-4}$. 
From measurements \citep{Planck2013abber} of the CMB aberration effect \citep{Challinor2002, Amendola2010, Kosowsky2010, Chluba2011ab} we have an independent confirmation for the dipole being mainly motion induced, so that this limit weakens as the primordial dipole becomes smaller.
Fitting for $G(x)$ and $Y^\ast(x)$ independently, we find $\Delta y_{\rm d}=\beta_{\rm pr}\,\Delta T^\ast_0/T_0=\pot{(1.1\pm0.6)}{-6}$, where $\beta_{\rm pr}$ is the primordial dipole amplitude. 
We assumed uncorrelated errors limited by noise (no foregrounds). This is unrealistic, since our motion with respect to any uniform background in the microwave bands will give rise to a dipole; however, a more sophisticated discussion is beyond the scope of this paper.
\citet{Fixsen1996} found $T_{\rm d}=(2.717\pm0.007)\rm K$ for the effective temperature of the dipole spectrum. This gives $ \Delta T^\ast_0/T_0\approx \pot{(-3.3\pm 5.0)}{-3}$ which is consistent with zero.
Additional constraints on the CMB anisotropy spectrum were discussed in \citet{Fixsen2003}. By improving these measurement, one could independently rule out changes of the TRR between recombination and today. Also, by combining measurements of the anisotropy spectrum with the dipole spectrum, one could furthermore constrain the primordial dipole contribution, if indeed the anisotropies indicate a change in the TRR.

\vspace{-0mm}
\section{Conclusion}
\label{sec:conclusions}
We discussed several effects related to changes of the TRR. In particular, we explained why at late times adiabatic photon production is extremely hard and a CMB spectral distortion is inevitably created unless very finely tuned models are considered. Thus, measurements of the TRR do generally not provide any improved constraint on scenarios with late photon production and instead should be reinterpreted as weak spectral distortion bounds for this case. Strong limits on the energy dependence of the photon production process can therefore be derived on a case-by-case basis using COBE/FIRAS. These constraint could improve dramatically in the future using a PIXIE-like concept.

We discussed ways to constrain inhomogeneous cosmologies using measurements of the CMB spectrum. Present bounds are already strong using COBE/FIRAS limits on the $y$-parameter \citep{Caldwell2008, Zibin2011, Zibin2011b, Bull2013, Caldwell2013}; however, pushing significantly deeper will be very hard even if future CMB distortion experiments could reach a level of $y\simeq 10^{-9}$. This is because we also expect a large $y$-distortion signal ($y\simeq 10^{-7}-10^{-6}$) from the reionization and structure formation era and possibly the pre-recombination era. 
We considered several effect to help disentangling these contributions to the signal. One consistency check is to probe the TRR and compare to model predictions. However, very sensitive measurements down to $\Delta T/T\simeq 10^{-4}-10^{-3}$ are required to make significant progress. Also, not all inhomogeneous models actually predict a large change in the monopole \citep[e.g.,][]{GarciaBellido2008}, and most observable signatures rather depend on the local dipole.
Constraints on a non-standard redshift scaling of the CMB quadrupole could be obtained with SZ polarization measurements; however, this only works for non-radially symmetric cosmologies, making this more contrived. Other multipoles of the CMB radiation field at higher redshift will be very hard to constrain individually, even in the future. Thus, fully ruling out inhomogeneous models down to the level of homogeneity expected in an FLRW cosmology will be extremely challenging.

Finally, we showed that applications of the thermal SZ effect to constrain the TRR are limited to scenarios with adiabatic photon production (which is hard to achieve) or changes to the TRR caused by scattering effects (see Sect.~\ref{sec:TRR_SZ}). However, clusters at $z>1$ are rare and the kSZ effect induced by inhomogeneous cosmologies between the cluster redshift and us greatly depends on the specific model, leaving it unclear how powerful this probe is. 
Changes in the TRR caused by gravitational effects (Sachs-Wolfe terms) do not change the shape of the SZ signal relative to the CMB blackbody part. In the latter case, tests based on the excitation states of interstellar species could still be used.

Non-gravitational changes to the TRR should furthermore give rise to an additional $y$-type dependence of the CMB temperature anisotropies. The measurements of Planck may already allow improving previous determination of the anisotropy spectrum based on WMAP and COBE/FIRAS \citep{Fixsen2003}. This could also be constrained in the future using a PIXIE-type experiment, providing another independent way to determine the primordial contribution to the CMB dipole and whether the TRR was affected non-gravitationally between recombination and today (see Sect.~\ref{sec:y_aniso}).

We close by mentioning scenarios with varying fundamental constants \citep[e.g.,][for models that are mediated by a scalar-photon coupling]{Avgoustidis2013, Barrow2013}. To avoid CMB spectral distortions limits, these models again need to follow the `adiabatic' condition for the changes of the photon energy and number density. In this case, a discussion similar to decaying vacuum scenarios and its difficulties applies. Thus, unless finely tuned, generally a CMB distortion should be created, which in principle allows us to place tight constraints on the energy dependence of the coupling using CMB spectroscopy. In particular, energy independent couplings seem to cause distortions. However, a more detailed analysis is beyond the scope of this work.

\small

\vspace{-5mm}
\section*{Acknowledgments}
The author cordially thanks Robert Caldwell, Dale Fixsen, Marc Kamionkowski, Carlos Martins and especially Dan Grin and James Zibin for very helpful comments and suggestions.
This work was supported by NSF Grant No. 0244990 and the John Templeton Foundation.


\bibliographystyle{mn2e}
\bibliography{Lit}

\end{document}